\documentclass[preprint,showpacs,preprintnumbers,amsmath,amssymb,superscriptaddress]{revtex4}
\usepackage{graphicx}
\usepackage{dcolumn}
\usepackage{bm}

\begin{document}
\preprint{preprint}

\title{NMR Study on the Vortex Slush Phase 
in Organic Superconductor $\kappa$-(BEDT-TTF)$_2$Cu(NCS)$_2$}
\author{M. Urano} 
\author{J. Tonishi} 
\author{H. Inoue} 
\author{T. Saito}
\author{T. Fujiwara}
\author{H. Chiku}
\author{A. Oosawa}
\affiliation{Department of Physics, Sophia University, 7-1 Kioi-cho, Chiyoda-ku, 
Tokyo 102-8554, Japan}
\author{T. Goto}
\email{gotoo-t@@sophia.ac.jp}
\affiliation{Department of Physics, Sophia University, 7-1 Kioi-cho, Chiyoda-ku, 
Tokyo 102-8554, Japan}
\affiliation{Institute for Material Research, Tohoku University, Sendai 980-8577}
\author{T. Suzuki}
\affiliation{Advanced Meson Science Laboratory, RIKEN 
(The Institute of Physical and Chemical Research)\\
2-1 Hirosawa, Wako, Saitama 351-0198}
\affiliation{Department of Physics, Sophia University, 7-1 Kioi-cho, Chiyoda-ku, 
Tokyo 102-8554, Japan}
\author{T. Sasaki}
\author{N. Kobayashi}
\author{S. Awaji}
\author{K. Watanabe}
\affiliation{Institute for Material Research, Tohoku University, Sendai 980-8577}

\date{\today}

\begin{abstract}
The vortex state in a single crystal of the layered organic superconductor 
$\kappa$-(BEDT-TTF$_2$Cu(NCS)$_2$, where BEDT-TTF (or ET) is 
bis(ethylenedithio)tetrathiafulvalene, was studied by $^1$H-NMR.  
Under a low field region around 0.75 T, 
the vortex glass-liquid transition was demonstrated by a diverging of the longitudinal 
nuclear spin relaxation rate and peak-broadening in spectra.  
Under a high field region near the upper critical field $H_{\rm c2}(0)\simeq$7 T,
the curvature of nuclear spin relaxation curves showed a drastic change at the temperature where the emergence of the quantum vortex slush state was reported.
The mechanism in this curvature change was discussed
in terms of the fluctuating field produced by fragments of vortex glass.
\end{abstract}

\maketitle

\section{Introduction}
Quantum fluctuation often brings a non-trivial physical state to atoms or solids
at low temperatures, represented by liquid $^4$He or spin-gap systems.  
The former persists to be liquid even at the absolute zero, and the latter shows a non-magnetic spin liquid state, 
where the N$\acute{\rm e}$el state is destructed by the quantum spin fluctuation.
Quantum melting of the flux lattice in superconductors has attracted much interest, 
because it is expected to bring about novel and exotic states of vortices.  
In superconductors with a quasi-two-dimensional structure, a short inter-plane spacing and a high critical temperature, 
the fluctuation is expected to have a large effect on the vortex state when the magnetic field is applied perpendicular to the conduction plane so that the intrinsic pinning has little effect on the motion of vortices\cite{Sasaki_slush,Sasaki_Hirr_finite_at_absolute_zero,Iketa_theory}.  
Although thermal fluctuation dominates the vortex state in the high temperature region near $T_{\rm c}$, Ikeda\cite{Iketa_theory} has shown that there is a crossover temperature $T_{\rm QF}$, below which a quantum fluctuation should be dominant, 
and that $T_{\rm QF}$ is proportional to
$H$ in the quasi-two-dimensional systems, 
so that the quantum melting phenomenon is expected to exist at a high field near $H_{\rm c2}$.  
In fact, Sasaki {\em et al.} made an intensive study on the layered organic superconductor
$\kappa$-(BEDT-TTF)$_2$Cu(NCS)$_2$, 
which is one of the most anisotropic systems in $\kappa$-type organic complexes, and showed that in the vicinity of $H_{\rm c2}$$\simeq$7 T,
a finite resistive state with non-ohmicity persists at low temperatures down to 100 mK.\cite{Sasaki_slush}
They conjectured from the results of transport measurements that 
the observed low resistance region corresponds to the slush state, where a short range order of vortices remains.   
The microscopic nature of such slush is still unexplored.  
The purpose of this study is to investigate the quantum vortex slush state microscopically by NMR.


\begin{figure}[h]
\includegraphics*[trim=1.2cm 9cm 17cm 1.5cm, clip, scale=0.75]{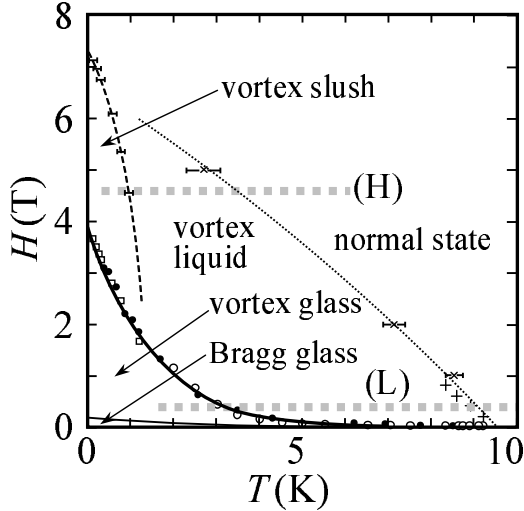}
\caption{
HT-phase diagram of $\kappa$-(BEDT-TTF)$_2$Cu(NCS)$_2$ reported by Sasaki {\em et al.}\cite{Sasaki_slush}   
and the experimental conditions of temperature range and fields performed in this paper,
denoted by two grey thick dotted lines.
Each curves show the boundary between the regions of the normal state, the vortex liquid phase, 
the vortex glass phase, the Bragg glass phase, and the quantum vortex slush state.   
Symbols of data points on curves are the same as in Ref. \cite{Sasaki_slush}.
}
\end{figure}

Figure 1 shows schematic HT-phase diagram of the vortex state in
$\kappa$-(BEDT-TTF)$_2$Cu(NCS)$_2$ reported in ref. \cite{Sasaki_slush},
where the phase boundary between the vortex glass and the vortex liquid, which
is usually referred as the irreversibility line $H_{\rm irr}$ intercepts 
the field around 4 T much lower than $H_{\rm c2}$.   
The slush state is located at low temperatures in the field region higher than 4 T.
So far, NMR studies on the title compound in powder samples were reported by Takahashi {\em et al.}\cite{Takahashi_CuNCS_NMR1,Takahashi_CuNCS_NMR2,Takahashi_CuNCS_NMR3,Takahashi_CuNCS_NMR4}
They found that under a low applied field around a Tesla, the longitudinal spin relaxation rate $T_1^{-1}$ shows a pronounced diverging behavior at temperatures far below $T_{\rm c}$.  
This anomaly, which coincides with the irreversibility line, is now understood to be due to the critical slowing down of magnetic field fluctuation at the vortex glass-liquid transition\cite{vortex_glass_Giamarchi}.  
Further examination on single crystals under a low field were reported by other groups.  
Van-quynh {\em et al.} observed that the $^1$H-NMR spectrum consists of a narrow peak and a broad peak overlapping each other, 
and argued that the latter comes from the part of the vortex lattice distorted by NMR excitation pulses\cite{CuNCS_NMR_vortex_deformated}.  
$\mu$-SR studies\cite{CuNCS_muSR,CuNCS_muSR2} report that a skewed Abrikosov pattern appears in the muon rotating spectra only under an extremely low transverse field (TF) of 25 Oe, corresponding to the Bragg-glass phase, and not at 400 Oe, corresponding to the vortex-glass phase.  
All these previous reports dealt with the low field region, 
where vortices solidify at low temperatures.  

In this paper, we report on a microscopic investigation of the vortex slush state located at high field
region above 4 T by $^1$H-NMR, which is a sensitive probe for the magnetic flux inside a superconductor.


\begin{figure}[h]
\includegraphics*[trim=1cm 12.3cm 2cm 0.3cm, clip, scale=0.52]{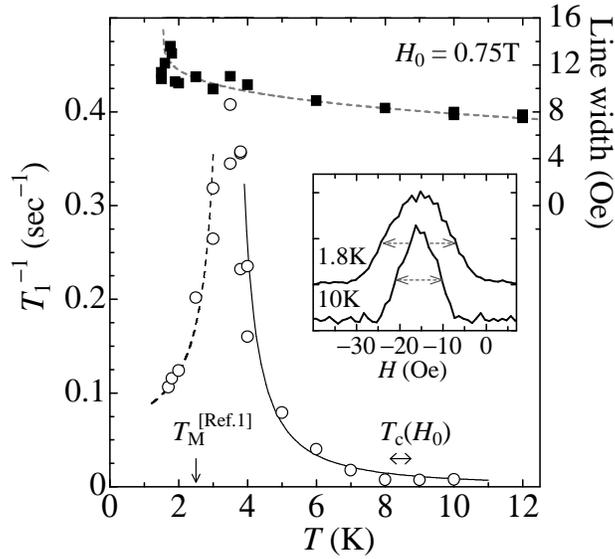}
\caption{
Temperature dependence of $T_1^{-1}$ and the peak width (FWHM) of $^1$H-NMR on the sample $\sharp$1 under the low field of 0.75 T, which was applied perpendicular to the conducting $bc$-plane.  
Solid and dashed curves are guides for the eyes.  The glass melting temperature $T_{M}$ and
the critical temperature $T_{\rm c}$ under $H_0$ is shown by solid arrows.
The inset shows typical spectra profiles.  
Horizontal dashed arrows indicate the definition of the line width (FWHM).
}
\end{figure}

\section{Experimental}
Single crystals of $\kappa$-(BEDT-TTF)$_2$Cu(NCS)$_2$ were synthesized with an electrical oxidation method.  
Two crystals $\sharp$1, 
$\sharp$2 with an approximate size of 3$\times$1.5$\times$0.1 mm were selected. 
$^1$H-NMR measurements were carried out separately on each two crystal 
by using a conventional $^3$He 
cryostat set in a 6T cryogen-free superconducting magnet with a field homogeneity of 
50 ppm in 10 mm dsv.  
In all the NMR measurements, the sample was field-cooled, with a slow cooling rate around 10-20 K/hour.  Spectra were obtained by superposing each Fourier-transformed spectra of the spin-echo beats taken under several magnetic fields with an interval of 10 Oe.  
The spin-lattice relaxation rate $T_1^{-1}$ in the low field region was determined 
by fitting the dependence of the echo amplitude on the repetition interval $\tau$
of measurements to the function 
\begin{equation}
I_0(1-e^{-\tau/T_1})
\end{equation}
where $I_0$ is the nuclear magnetization at thermal equilibrium.
In the high field region, where the slush state emerges, 
$T_1$ exceeds a few hundred minutes, making it difficult to determine its absolute value.  Instead, we have traced the initial curvature of the relaxation curves and investigated its temperature dependence.

\section{Results and Discussion}
Figure 2 shows the temperature dependence of $T_1^{-1}$ at a low field of 0.75 T, which corresponds
to the line (L) in Fig. 1.
The field was applied perpendicular to the conducting $bc$-plane.   
One can see that $T_1^{-1}$ diverges around 3 K, which is much lower than $T_{\rm c}(H = 0.75 {\rm T})\simeq$ 8.5 K.  
This temperature coincides with the vortex liquid-glass transition reported by NMR on powder samples \cite{Takahashi_CuNCS_NMR1,Takahashi_CuNCS_NMR2,Takahashi_CuNCS_NMR3,Takahashi_CuNCS_NMR4} and by magnetization measurements \cite{Sasaki_slush,Sasaki_Hirr_finite_at_absolute_zero,Sasaki_and_Nishizaki_Hirr_at_low_field} and we can safely assign the divergence in $T_1^{-1}$ to the vortex glass melting, 
where the fluctuation in the magnetic field produced by vortex motion shows a critical slowing down.  
Generally, the nuclear spin-lattice relaxation is driven by a Fourier component of the Lamor frequency, that is, around 100-200 MHz, in a fluctuating magnetic field.  
This slow component is prominently enhanced in the vicinity of second order phase transitions such as glass melting\cite{Takahashi_CuNCS_NMR1,Takahashi_CuNCS_NMR2,Takahashi_CuNCS_NMR3,Takahashi_CuNCS_NMR4}.

The temperature dependence of the width of the spectral peak defined as the full width at half maximum (FWHM) is also shown in Fig. 2.  
With decreasing temperature, the width shows a slight increase from 8 to 11 Oe, and an abrupt increase to 14 Oe at the temperature slightly lower than that where $T_1^{-1}$ diverges.  
This difference comes from the fact that the $T_1^{-1}$ diverges when the dominant frequency component
 of the field produced by vortex coincides with the Lamor frequency while the spectrum broadens when
the field becomes nearly static.

The observed spectra shown in the inset always consist of a symmetric peak and bear neither skewness nor splitting in the entire temperature range, in contrast to the previous report\cite{CuNCS_NMR_vortex_deformated}.  
Generally, in the presence of a rigid Abrikosov flux lattice with translational symmetry, the internal field distribution must show a characteristic skewed pattern, depending on the lattice symmetry \cite{Redfield_pattern}.  
The absence of the skewness in the observed spectra indicates that the vortex pattern loses its translational symmetry within each plane.  This is consistent with the fact that the system resides in the vortex glass state under an applied field of 0.75 T.

\begin{figure}
\includegraphics*[trim=1cm 6.5cm 1cm 2cm, clip, scale=0.5]{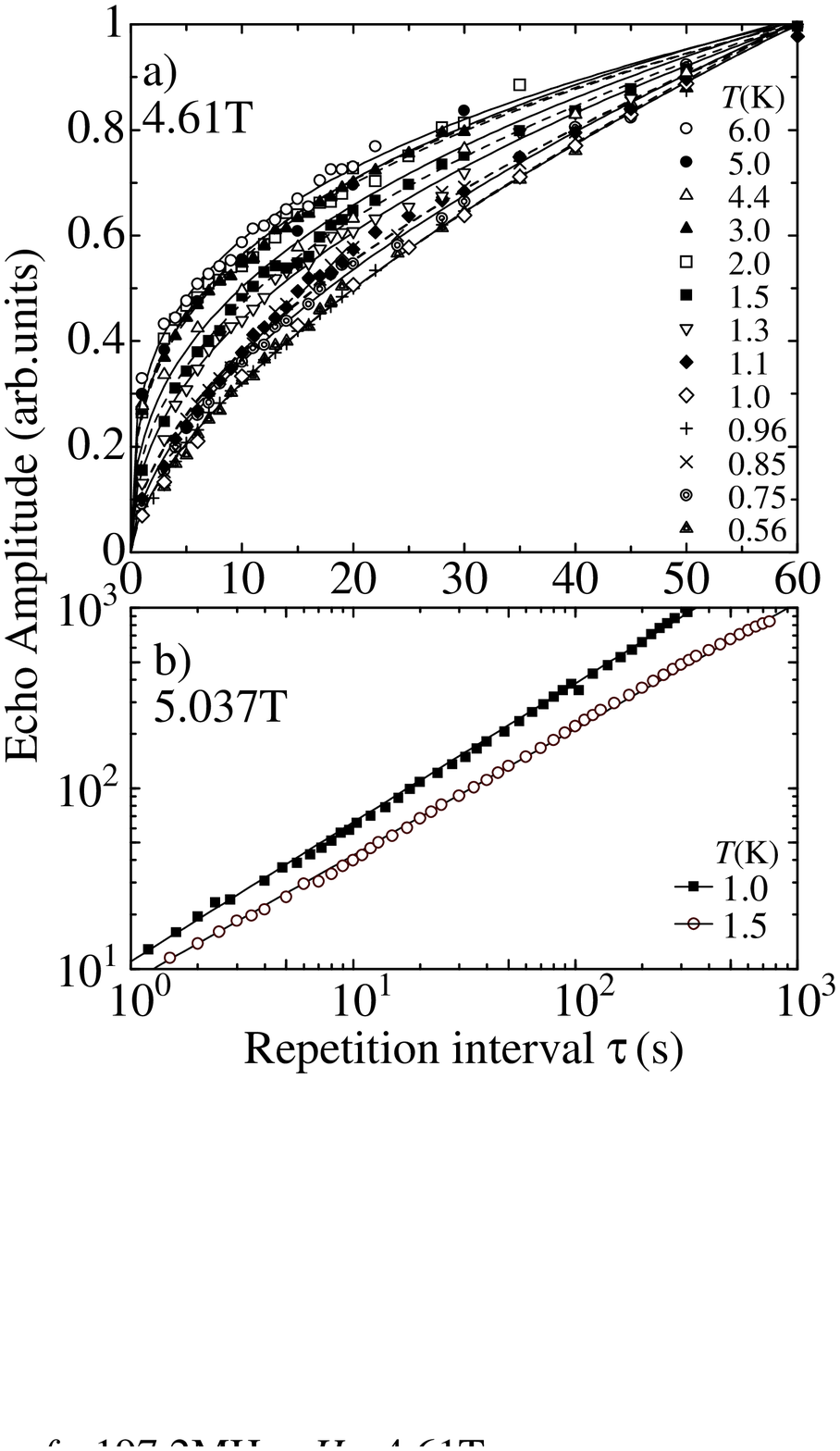}
\caption{
Typical relaxation curves at high field around 5T, a) for initial part within 60 s for the sample $\sharp$1, and b) within $10^3$ s of the sample $\sharp$2.   
Solid and dashed curves indicate the power law function.
}
\end{figure}

Next, we proceed to the high field data, the field and temperature region of which corresponds to 
the line (H) in Fig. 1.
The spectrum consists of a single peak also in this field region, and its width of 14(5) Oe does not show any temperature dependence between 15 and 0.5 K.  Figure 3 shows the initial part of typical relaxation curves at various temperatures.  Under a high field of around 4-5 T, the relaxation time is drastically elongated to be far above $10^4$ s in the measured temperature region below 6 K.  
Therefore, the absolute value of $T_1$ and hence its temperature dependence could not be precisely determined in the allowable experiment time.  
The initial part of the relaxation curves, at least from zero to 2$\times$10$^3$ s, 
exactly follows the power law of time as shown by the log-log plot in Fig. 3(b).  
It took over a day to obtain one of the relaxation curves in the figure.  
As lowering the temperature, the relaxation curve loses its curvature and tends to become a straight line as can be seen in Fig. 3 (a).

We focus attention on the power-law index obtained by fitting the relaxation curves to the 
function $\tau^{\alpha}$, where $\alpha$ is a constant.  
Figure 4 shows the temperature dependence of $\alpha$, 
which stays constant in the high temperature region, shows a kink at around 1.5 K, and increases steeply to unity at low temperatures.  
This kink temperature is nearly the same as the onset temperature of the slush state, where the resistance starts to decrease steeply toward the residual low value \cite{Sasaki_slush}.  
The non-unity value in $\alpha$ and its anomalous temperature dependence are characteristic of the present system, whereas the other $\kappa$-type organic superconductors, such as 
$\kappa$-(BEDT-TTF)$_2$Cu[N(CN)$_2$]Br, 
which has less two-dimensionality and hence does not show any quantum vortex slush state\cite{Br_NMR_T1,Br_NMR_T1_Mayaffre}.  
These facts indicate that the resistance drop observed by Sasaki {\em et al}. corresponds to a microscopic change in the vortex state.  
The sharpness in the kink of the power index $\alpha$ suggests that the change is not a crossover but corresponds to a remnant of the vortex glass-liquid phase transition, 
which has been obscured by quantum fluctuation and displaced to lower temperatures\cite{Sasaki_slush}.

Generally, power-law-type relaxation curves with non-unity $\alpha$ come from the distribution in $T_1$, and are often seen in disordered systems, such as non-magnetic systems containing dilute relaxation centers \cite{stretched_exponential}.  
In these systems, each nuclear spins feel the field fluctuation with a different amplitude depending on the distance from the nearest neighboring relaxation center, and hence has a different $T_1^{-1}$, which is proportional to the squared fluctuation amplitude.  
Since the present system is in the clean limit and free from magnetic impurities, we consider that the relaxation center has its origin in, for example, crystal dislocations which has an extremely low spatial concentration.  
The slight dependence of $\alpha$ in the liquid state on the sample $\sharp$1 and $\sharp$2 may be rooted in the difference in their concentration in each crystal.  
The vortices in the liquid state contribute little to the nuclear spin relaxation, because they are considered to have a field fluctuation spectrum weighted at the frequency much higher than the NMR Lamor frequency.   
Upon entering the slush state from the liquid state, there appear fragments of vortex glass in the liquid.   
These fragments, which waft slowly in the liquid is expected to drive the nuclear relaxation.  As the concentration of the fragments increases with decreasing temperature, 
all the proton nuclear spins feel the field fluctuation with nearly the same amplitude, and will show a simple exponential recovery, that corresponds to $\alpha$=1.

\begin{figure}
\includegraphics*[trim=2cm 12.8cm 1cm 0.7cm, clip, scale=0.52]{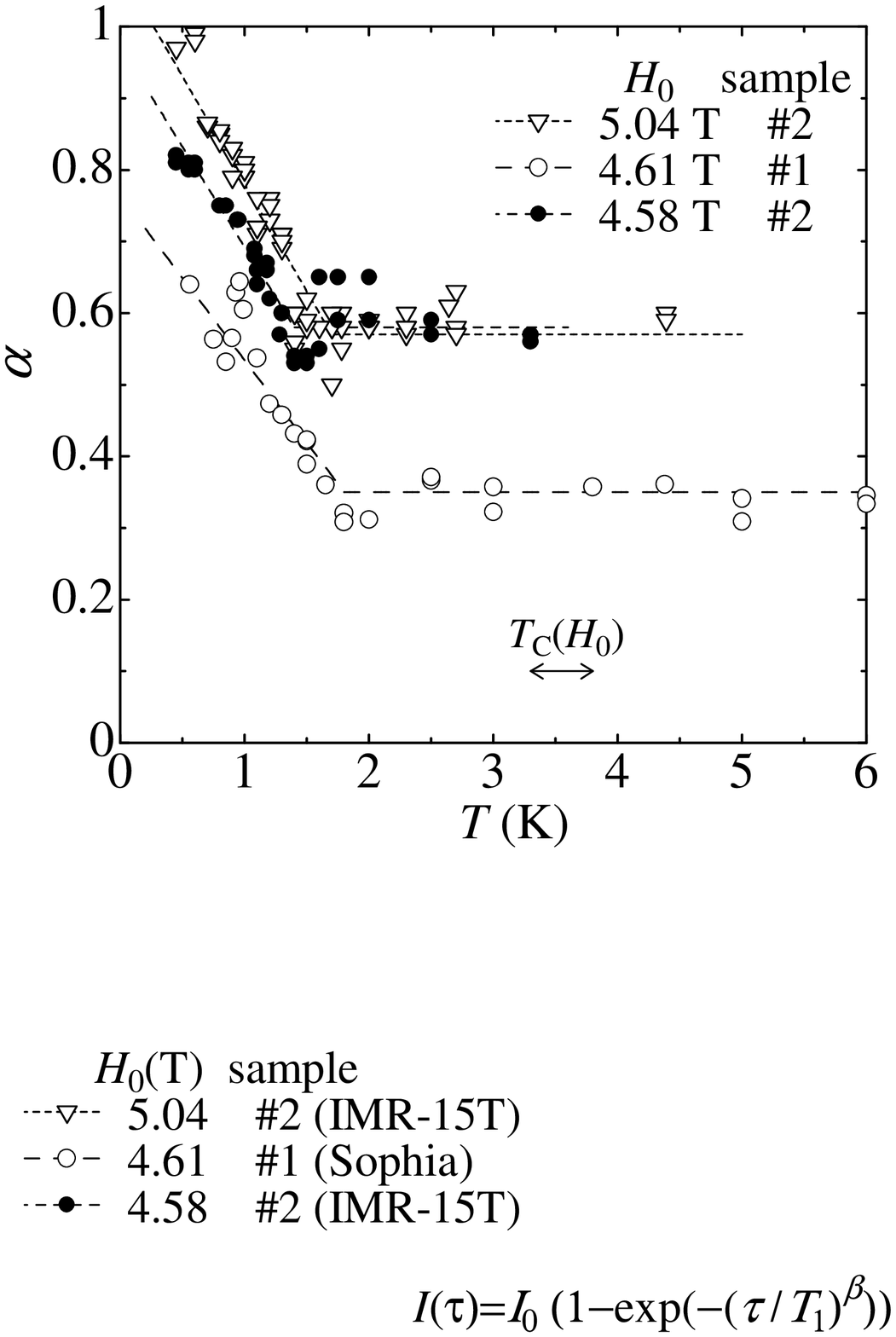}
\caption{
Temperature dependence of $\alpha$, the power law index of the relaxation curve, 
the time evolution of which obeys to the power law $\tau^\alpha$.  
The dashed curves are guides for the eyes.  
}
\end{figure}

Finally, we comment on the effect of the nuclear spin diffusion on the relaxation.  
The appearance of power-law-type relaxation curves indicates that the fast nuclear spin diffusion is absent in this system.  In the presence of fast nuclear spin diffusion, 
the normal electrons in the vortex core must dominate the nuclear spin relaxation, 
which is expected to be proportional to the applied field, 
as is often observed in high-$T_{\rm c}$ cuprates\cite{Spin_diffusion_normal_core_NMR_Hg1223,Spin_diffusion_normal_core_NMR_YBCO}.  
Note that in this system, on the contrary, $T_1^{-1}$ is suppressed in the higher field.  
We can estimate the mean distance between dislocations that act as relaxation centers, 
from the condition that the nuclear spin diffusion does not propagate from one relaxation center to its nearest neighboring center within a period of time 
$\tau \simeq 2\times 10^3$ s in which power-law-type relaxation is observed.  
The lower limit of the mean distance is given as  
\begin{equation}
\ell\simeq\sqrt{D\tau}\simeq\sqrt{(L^2/T_2)\tau}\simeq 1\times 10^4 \mbox{ - } 3\times10^4{\rm \AA}
\end{equation}
where  $L\simeq 5 {\rm \AA}$ is the mean distance between BEDT-TTF molecules, 
$T_2\simeq$ 30-100 $\mu$s, 
the nuclear spin-spin relaxation rate between the two proton nuclei belonging to adjacent two BEDT-molecules.  
This estimation is consistent with the report by Shubnikov-de Haas effect\cite{Sasaki_SdH_CuNCS}, 
supporting the validity of our interpretation of the power-law-type relaxation.

\section{Summary}
We have investigated NMR spectra and relaxations 
in the quasi-two-dimensional superconductor $\kappa$-(BEDT-TTF)$_2$Cu(NCS)$_2$.
The vortex glass-liquid transition under a low field of around 0.75 T was confirmed from the diverging $T_1^{-1}$ and the spectral broadening.  
Under high field around 4.6 - 5 T, the temperature dependence of the power index $\alpha$ of the nuclear spin relaxation curve showed a kink at the vortex liquid-slush transition temperature, below which, $\alpha$ approached unity.  
This behavior of $\alpha$ is understood in terms of an enhancement of a short range order of vortices on entering into the slush state, indicating the first microscopic observation of the vortex slush. 

\section*{Acknowledgments}
This work was supported in part by a Grant-in-Aid for Scientific Research on the Priority Area ``High Magnetic Field Spin Science in 100T'' from MEXT, and by the Kurata Memorial Hitachi Science and Technology Foundation and Saneyoshi Scholarship Foundation.
A part of this study was performed at 
the High Field Laboratory for Superconducting Materials, Institute for Materials Research, Tohoku University.


\end{document}